\documentclass[12pt]{article}
  \usepackage{amsfonts}
  \usepackage{amsmath}
\usepackage{amssymb}
\usepackage{amscd}
\usepackage[dvips]{graphicx}
\begin{document}
~~
\bigskip
\bigskip
\begin{center}
{\Large {\bf{{{Twist deformations of Newton-Hooke Hopf algebras}}}}}
\end{center}
\bigskip
\bigskip
\bigskip
\begin{center}
{{\large ${\rm {Marcin\;Daszkiewicz}}$ }}
\end{center}
\bigskip
\begin{center}
{ {{{Institute of Theoretical Physics\\ University of Wroc{\l}aw pl.
Maxa Borna 9, 50-206 Wroc{\l}aw, Poland\\ e-mail:
marcin@ift.uni.wroc.pl}}}}
\end{center}
\bigskip
\bigskip
\bigskip
\bigskip
\bigskip
\bigskip
\bigskip
\bigskip
\begin{abstract}
We construct five new quantum Newton-Hooke Hopf algebras with the
use of Abelian twist procedure. Further we  demonstrate that the
corresponding deformed space-times with quantum space and classical
time are periodic or expanding in time.
\end{abstract}
\bigskip
\bigskip
\bigskip
\bigskip
\bigskip
\bigskip
\bigskip
\bigskip
\bigskip
 \eject
\section{{{Introduction}}}

The two  Newton-Hooke cosmological algebras ${ NH}_{\pm}$ (acting on
commutative space-time) were introduced in \cite{bacry} with the use
of nonrelativistic contraction of the de Sitter and anti-de Sitter
groups respectively (see also \cite{contra1}-\cite{contra3}).
The characteristic 
time scale $\tau$ present in both  algebras can be interpreted in
terms of the inverse of Hubble's constant
 for the expanding universe $({ NH}_{+})$ or associated
to the "period" for the oscillating case $({ NH}_{-})$. Obviously,
for time parameter $\tau$ approaching infinity one  gets the
Galilean symmetry acting on  the standard (flat) nonrelativistic
space-time. The Newton-Hooke symmetries has found an application in
nonrelativistic cosmology \cite{cosmo1}-\cite{cosmo3} as well as in
M- and string theory \cite{string}.

In this article we discuss the role which can be played by
Newton-Hooke symmetries in a context of noncommutative
geometry\footnote{Particulary, we ask about the impact  of $\tau$
parameter  on the form of quantum space.}. The suggestion to use
noncommutative coordinates goes back to Heisenberg and was
formalized by Snyder in \cite{snyder}. Recently, there were also
found arguments based on Quantum Gravity \cite{grav1}, \cite{grav2}
and String Theory models \cite{recent}, \cite{string1} indicating
that space-time at Planck scale  should be noncommutative, i.e. it
should  have a quantum nature.

In our treatment  we consider  the Abelian (Reshetikhin) twist
deformations (see \cite{twist1}, \cite{twist2}) of the Newton-Hooke
Hopf algebras $\,{\mathcal U}_0({ NH}_{\pm})$\footnote{The
Newton-Hooke Hopf algebras $\,{\mathcal U}_0({ NH}_{\pm})$ can be
obtained by nonrelativistic contraction of the classical de Sitter
and anti-de Sitter quantum groups respectively (see e.g.
\cite{balle1}, \cite{balle2}). They are given by algebraic
commutation relations for ${ NH}_{\pm}$ groups supplemented by the
trivial coproduct sector $\Delta_0(a) = a\otimes 1+ 1\otimes a$.}.
In such a way we get five new quantum groups providing
noncommutative $NH$ (Newton-Hooke) space-times. It should be noted,
however, that similar  investigation has been already performed in
the case of standard relativistic (Poincare) and nonrelativistic
(Galilei) groups. Consequently, in accordance with the
Hopf-algebraic classification of all deformations of mentioned
symmetries, one can distinguish three
kinds of quantum spaces \cite{class1}, \cite{class2}:\\
\\
{ \bf 1)} Canonical ($\theta^{\mu\nu}$-deformed) space-time
\begin{equation}
[\;{ x}_{\mu},{ x}_{\nu}\;] = i\theta_{\mu\nu}\;\;\;;\;\;\;
\theta_{\mu\nu} = {\rm const}\;, \label{noncomm}
\end{equation}
introduced in  \cite{oeckl}, \cite{chi} in the case of Poincare
quantum
group, and in \cite{dasz1}, \cite{dasz2} for its Galilean counterpart.\\
\\
{ \bf 2)} Lie-algebraic modification of classical space
\begin{equation}
[\;{ x}_{\mu},{ x}_{\nu}\;] = i\theta_{\mu\nu}^{\rho}{ x}_{\rho}\;,
\label{noncomm1}
\end{equation}
with  particularly chosen coefficients $\theta_{\mu\nu}^{\rho}$
being constants. This type of noncommutativity has been  represented
by $\kappa$-Poincare \cite{kappaP} and  $\kappa$-Galilei
\cite{kappaG} as well as by twisted relativistic \cite{lie2} (see
also \cite{lie1}) and  nonrelativistic \cite{dasz1},
\cite{dasz2} symmetries respectively. \\
\\
{ \bf 3)} Quadratic deformation of Minkowski and Galilei  space
\begin{equation}
[\;{ x}_{\mu},{ x}_{\nu}\;] = i\theta_{\mu\nu}^{\rho\tau}{
x}_{\rho}{ x}_{\tau}\;, \label{noncomm2}
\end{equation}
with coefficients $\theta_{\mu\nu}^{\rho\tau}$ being constants. This
kind of deformation has been proposed in \cite{qdef}, \cite{paolo},
\cite{lie2}
 at relativistic and in \cite{dasz2} at  nonrelativistic level.\\
\\
In this article we demonstrate that in the case  of Newton-Hooke
Hopf algebras $\,{\mathcal U}_0({ NH}_{\pm})$ the twist deformation
provides the new  space-time noncommutativity which is expanding
($\,{\mathcal U}_0({ NH}_{+})$) or periodic ($\,{\mathcal U}_0({
NH}_{-})$) in time, i.e. it takes the form\footnote{$x_0 = ct$.},
\footnote{The mentioned space-times are defined as the Hopf modules
of twisted Newton-Hooke Hopf algebras respectively (see e.g.
\cite{bloch}, \cite{wess}, \cite{chi}).}
\begin{equation}
[\;t,{ x}_{i}\;] = 0\;\;\;,\;\;\; [\;{ x}_{i},{ x}_{j}\;] = 
f_{\pm}\left(\frac{t}{\tau}\right)\theta_{ij}(x)
\;, \label{nhspace}
\end{equation}
with time-dependent  functions
$$f_+\left(\frac{t}{\tau}\right) =
f\left(\sinh\left(\frac{t}{\tau}\right),\cosh\left(\frac{t}{\tau}\right)\right)\;\;\;,\;\;\;
f_-\left(\frac{t}{\tau}\right) =
f\left(\sin\left(\frac{t}{\tau}\right),\cos\left(\frac{t}{\tau}\right)\right)\;,$$
and $\theta_{ij}(x) \sim \theta_{ij} = {\rm const}$ or
$\theta_{ij}(x) \sim \theta_{ij}^{k}x_k$. In such a way we indicate
the impact of time scale $\tau$ on the structure of quantum space,
i.e. we show that the time parameter  $\tau$  is responsible for
oscillation or expansion of space-time noncommutativity. Of course,
for time scale $\tau$ running to infinity we reproduce the canonical
(\ref{noncomm}), Lie-algebraic  (\ref{noncomm1}) and quadratic
(\ref{noncomm2}) twisted Galilei space-times provided in
\cite{dasz1} and \cite{dasz2}.

It should be noted that similar investigations have been already
performed in \cite{balle2}, \cite{x2} at relativistic level.
Particulary, it has been shown that one can get from q-deformed
(anti-)de Sitter space-time (containing cosmological constant) the
Lie-algebraically $\kappa$-deformed Minkowski space. However, it
should be also mentioned, that considered in \cite{balle2} and
\cite{x2} (anti-)de Sitter space-times have been obtained in
different framework in comparision with the technique used in this
paper, as the translation sector of corresponding dual Hopf
structure.

Finally, let us note that the  motivations for  present studies are
manyfold. First of all, such investigations are interesting because
they provide five new explicit quantum groups and the corresponding
noncommutative space-times. Besides, it should be noted, that the
provided Newton-Hooke Hopf algebras permit to construct the
corresponding phase-spaces in the framework of so-called Heisenberg
double procedure  \cite{twist1}. Consequently, it permits us to
discuss of Heisenberg uncertainty principle \cite{phamelia}
associated with such deformed quantum space-times. Finally,  one can
consider corresponding classical and quantum nonrelativistic
particle models. Such a construction has been already presented (see
\cite{walczyk1}, \cite{walczyk2}) in the case of  classical particle
moving in external constant force on the  space-times
(\ref{noncomm})-(\ref{noncomm2}). The studies of deformations
(\ref{nhspace}) in a context of basic dynamical models seems quite
interesting and  is postponed for further investigations\footnote{It
seems quite interesting to check, for example, the possible
connection of such deformed Newton-Hooke symmetries with Modified
Newtonian Dynamics (MOND) model \cite{mond}.}.

The paper is organized as follows. In second Section the new twisted
Newton-Hooke Hopf algebras are provided. The corresponding deformed
space-times with quantum space and classical time are derived in
Section 3. The final remarks are presented in the last Section.

\section{{{Twist deformed Newton-Hooke Hopf algebras}}}

In this Section we provide five twisted Newton-Hooke Hopf algebras.
All of them are described by the following (Abelian) classical
r-matrices\footnote{The symbols $M_{ab}$, $K_a$, $P_a$ and $H$
denote rotations, boosts  and space-time translation generators
respectively.}
\begin{eqnarray}
r_{\alpha_1} &=&  \frac{1}{2}{\alpha_1^{kl}} P_k \wedge
P_l\;\;\;\;\;\;\;\,
[\;\alpha_1^{kl} = -\alpha_1^{lk}\;]\;, \label{macierze0}\\
&~~&\cr r_{\alpha_2} &=& \frac{1}{2}\alpha_2^{kl} K_k \wedge
P_l\;\;\;\;\;\;\; [\;\alpha_2^{kl} = -\alpha_2^{lk}\;]\;,\\ &~~&\cr
r_{\alpha_3} &=& \frac{1}{2}{\alpha_3^{kl}} K_k \wedge
K_l\;\;\;\;\;\;\, [\;\alpha_3^{kl} = -\alpha_3^{lk}\;]\;,
\label{macierze}\\
&~~&\cr r_{\alpha_4} &=&  \alpha_4 K_m \wedge M_{kl}\;\;\; [\;m,k,l
- {\rm fixed},\;\;m \neq k,l\;]\;,\\ &~~&\cr r_{\alpha_5} &=&
\alpha_5 P_m \wedge M_{kl}\;\;\; [\;m,k,l - {\rm fixed},\;\;m \neq
k,l\;]\;,\label{macierze1}
\end{eqnarray}
satisfying the classical Yang-Baxter equation (CYBE)
\begin{equation}
[[\;r_{\cdot},r_{\cdot}\;]] = [\;r_{\cdot 12},r_{\cdot13} + r_{\cdot
23}\;] + [\;r_{\cdot 13}, r_{\cdot 23}\;] = 0\;, \label{cybe}
\end{equation}
where  the  symbol $[[\;\cdot,\cdot\;]]$ denotes the Schouten
bracket, $a\wedge b = a\otimes b - b\otimes a$, and for $r =
\sum_{i}a_i\otimes b_i$
$$r_{ 12} = \sum_{i}a_i\otimes b_i\otimes 1\;\;,\;\;r_{ 13} = \sum_{i}a_i\otimes 1\otimes b_i\;\;,\;\;
r_{ 23} = \sum_{i}1\otimes a_i\otimes b_i\;.$$


In accordance with twist procedure \cite{twist1}, \cite{twist2} the
algebraic sector of all algebras remains classical
\begin{eqnarray}
&&\left[\, M_{ab},M_{cd}\,\right] =i\left( \delta
_{ad}\,M_{bc}-\delta _{bd}\,M_{ac}+\delta _{bc}M_{ad}-\delta
_{ac}M_{bd}\right)\;\; \;, \;\;\; \left[\, H,P_a\,\right] =\pm
\frac{i}{\tau^2}K_a
 \;,  \notag \\
&~~&  \cr &&\left[\, M_{ab},K_{c}\,\right] =i\left( \delta
_{bc}\,K_a-\delta _{ac}\,K_b\right)\;\; \;, \;\;\;\left[
\,M_{ab},P_{c }\,\right] =i\left( \delta _{b c }\,P_{a }-\delta _{ac
}\,P_{b }\right) \;, \label{nnnga}
\\
&~~&  \cr &&\left[ \,M_{ab},H\,\right] =\left[ \,K_a,K_b\,\right] =
\left[ \,K_a,P_{b }\,\right] =0\;\;\;,\;\;\;\left[ \,K_a,H\,\right]
=-iP_a\;\;\;,\;\;\;\left[ \,P_{a },P_{b }\,\right] = 0\;,\nonumber
\end{eqnarray}
while the (twisted) coproducts and antipodes  take the form
\begin{equation}
\Delta _{0}(a) \to \Delta _{\cdot }(a) = \mathcal{F}_{\cdot }\circ
\,\Delta _{0}(a)\,\circ \mathcal{F}_{\cdot }^{-1}\;\;\;,\;\;\;
S_{\cdot}(a) =u_{\cdot }\,S_{0}(a)\,u^{-1}_{\cdot }\;,\label{fs}
\end{equation}
with $\Delta _{0}(a) = a \otimes 1 + 1 \otimes a$, $S_0(a) = -a$ and
$u_{\cdot }=\sum f_{(1)}S_0(f_{(2)})$ (we use Sweedler's notation
$\mathcal{F}_{\cdot }=\sum f_{(1)}\otimes f_{(2)}$). Present in the
commutation relations (\ref{nnnga}) parameter $\tau$ denotes the
characteristic for Newton-Hooke algebra 
time scale. In the limit $\tau \to \infty$ we get from $\,{\mathcal
U}_{0}({ NH}_{\pm})$ algebras the standard Galilei Hopf structure
$\,{\mathcal U}_0({\mathcal G})$. It should be also noted, that one
can  pass from $\,{\mathcal U}_{0}({ NH}_{\pm})$ groups to
$\,{\mathcal U}_{0}({ NH}_{\mp})$ algebras  by the multiplication of
$H$ and $P_a$ generators by
the imaginary element $i=\sqrt{-1}$.\\
The  twist factors $\mathcal{F}_{\cdot } \in {\mathcal U}_{0}({
NH}_{\pm}) \otimes {\mathcal U}_{0}({ NH}_{\pm})$ satisfy  the
classical cocycle condition
\begin{equation}
{\mathcal F}_{{\cdot }12} \cdot(\Delta_{0} \otimes 1) ~{\cal
F}_{\cdot } = {\mathcal F}_{{\cdot }23} \cdot(1\otimes \Delta_{0})
~{\mathcal F}_{{\cdot }}\;, \label{cocyclef}
\end{equation}
and the normalization condition
\begin{equation}
(\epsilon \otimes 1)~{\cal F}_{{\cdot }} = (1 \otimes
\epsilon)~{\cal F}_{{\cdot }} = 1\;, \label{normalizationhh}
\end{equation}
with ${\cal F}_{{\cdot }12} = {\cal F}_{{\cdot }}\otimes 1$ and
${\cal F}_{{\cdot }23} = 1 \otimes {\cal F}_{{\cdot }}$. They look
as follows
\begin{eqnarray}
{\cal F}_{\alpha_1} &=&  \exp \left(\frac{i}{2}{\alpha_1^{kl}} P_k
\wedge P_l \right)\;\;\;\;\;\;\;\, [\;\alpha_1^{kl} =
-\alpha_1^{lk}\;]\;, \label{factory0}\\ &~~&\cr {\cal F}_{\alpha_2}
&=& \exp \left(\frac{i}{2}{\alpha_2^{kl}} K_k \wedge
P_l\right)\;\;\;\;\;\;\; [\;\alpha_2^{kl} = -\alpha_2^{lk}\;]\;,\\
&~~&\cr {\cal F}_{\alpha_3} &=& \exp
\left(\frac{i}{2}{\alpha_3^{kl}} K_k \wedge K_l\right)\;\;\;\;\;\;\,
[\;\alpha_3^{kl} = -\alpha_3^{lk}\;]\;,
\label{factory}\\
&~~&\cr {\cal F}_{\alpha_4} &=&  \exp \left(i\alpha_4 K_m \wedge
M_{kl}\right)\;\;\; [\;m,k,l - {\rm fixed},\;\;m \neq k,l\;]\;,\\
&~~&\cr {\cal F}_{\alpha_5} &=& \exp \left(i\alpha_5 P_m \wedge
M_{kl} \right)\;\;\; [\;m,k,l - {\rm fixed},\;\;m \neq
k,l\;]\;.\label{factory1}
\end{eqnarray}
Hence, in accordance with the formula (\ref{fs}), the antipodes
remain classical while the corresponding twisted  coproducts take
the form
\begin{eqnarray}
&&\Delta_{\alpha_1}(M_{ab}) =\Delta_0(M_{ab})-\alpha_{1
}^{kl}[\,(\delta_{k a}P_{b }-\delta_{k b }\,P_{a})\otimes P_{l
}+P_{k}\otimes (\delta_{l
a}P_{b}-\delta_{l b}P_{a})\,]\;,\label{coproduct0}\\
&~~& \cr &&\Delta_{\alpha_1}(K_a) =\Delta_0(K_a)\;\;\;,\;\;\;
\Delta_{\alpha_1}(P_a) = \Delta_0(P_a)
\;\;\;,\;\;\;\Delta_{\alpha_1}(H) = \Delta_0(H) \pm
\frac{\alpha_1^{kl}}{\tau^2}K_k \wedge P_l
\;\;\;\;\;\;\;\;\label{coproduct1}
\end{eqnarray}
in the case of first deformation
\begin{eqnarray}
&&\Delta_{\alpha_2}(M_{ab}) =\Delta_0(M_{ab})-
\frac{i}{2}\alpha_2^{kl}\left[\,M_{ab},K_k\,\right]\wedge P_l
-\frac{\alpha_2^{kl}}{2}K_k \wedge(\delta_{al}P_b -\delta_{bl}P_a)
\;,\label{coproduct00} \\&~~& \cr &&\Delta_{\alpha_2}(K_a)
=\Delta_0(K_a)\;\;\;,\;\;\; \Delta_{\alpha_2}(P_a) =
\Delta_0(P_a)\;,\label{coproduct2}\\
&~~&  \cr  &&\Delta_{\alpha_2}(H) = \Delta_0(H) \pm
\frac{\alpha_2^{kl}}{2\tau^2}K_k \wedge K_l +
\frac{\alpha_2^{kl}}{2}P_k \wedge P_l\;,\label{coproduct2a}
\end{eqnarray}
for the second Hopf structure
\begin{eqnarray}
&&\Delta_{\alpha_3}(M_{ab}) = \Delta_0(M_{ab})-\alpha_{3
}^{kl}[\,(\delta_{k a}K_{b }-\delta_{k b }\,K_{a})\otimes K_{l
}+K_{k}\otimes (\delta_{l a}K_{b}-\delta_{l b}K_{a})\,]
 \;,\label{coproduct30} \\&~~&  \cr && \Delta_{\alpha_3}(K_a) =
\Delta_0(K_a)\;\;\;,\;\;\;\Delta_{\alpha_3}(P_a) =
\Delta_0(P_a)\;\;\;,\;\;\;\Delta_{\alpha_3}(H) =\Delta_0(H) +
\alpha_3^{kl}P_l\wedge
K_k\;\;\;\;\;\;\;\;\label{coproduct3a}
\end{eqnarray}
for third Newton-Hopf algebra
\begin{eqnarray}
&&\Delta_{\alpha_4}(M_{ab}) = \Delta_0(M_{ab}) +M_{k l }\wedge
\alpha_4 \left(\delta_{a
m }K_b-\delta_{b m }K_a\right)\nonumber\\
&&~~~~~~~~~~~~~~~~~~~~~~~~~~~~+i\left[\,M_{ab},M_{kl}\,\right]\wedge
\sin(\alpha_4 K_m ) \notag \\
&&~~~~~~~~~~~~~~~~~~~~~~~~+\left[\,\left[\, M_{ab},M_{k l
}\,\right],M_{k l }\,\right]\perp
(\cos(\alpha_4  K_m  )-1) \label{coproduct40} \\
&&~~~~~~~~~~~~~~~~~~~~~~~~+ M_{k l }\sin(\alpha_4 K_m )\perp
\alpha_4 \left(\psi_m K_k -\chi_m K_l \right) \nonumber  \\
&&~~~~~~~~~~~~~~~~~~~~~~~~+ \alpha_4 \left(\psi_m K_l +\chi_m K_k
\right)\wedge M_{k l
}(\cos(\alpha_4 K_m )-1)\;, \nonumber\\
\cr &~~&  \cr &&\Delta_{\alpha_4}(K_a) = \Delta_0(K_a)
+i\left[\,K_a,M_{k l }\,\right]\wedge \sin(\alpha_4 K_m )
\label{coproduct4}\\
&&~~~~~~~~~~~~~~~~~~~~~~~~~~~~~~~~~~~+\left[\,\left[%
K_a,M_{k l }\,\right],M_{k l }\,\right]\perp (\cos(\alpha_4 K_m
)-1)\;,\nonumber\\
&~~&  \cr &&\Delta_{\alpha_4}(P_a) = \Delta _0(P_a)+\sin( \alpha_4
K_m )\wedge
\left(\delta_{k a}P_l -\delta_{l a}P_k \right)\\
&&~~~~~~~~~~~~~~~~~~~~~~~~~~~~~~~~~~~+(\cos(\alpha_4  K_m )-1)\perp
\left(\delta_{k a}P_k +\delta_{l a}P_l \right)\;, \cr &~~&  \cr
&&\Delta_{\alpha_4}(H) = \Delta_0(H) - \alpha_4 M_{kl}\wedge
P_m\;,\nonumber
\end{eqnarray}
in the case of forth deformation, and finally,
\begin{eqnarray}
&&\Delta_{\alpha_5}(M_{ab}) = \Delta_0(M_{ab}) +M_{k l }\wedge
\alpha_5 \left(\delta_{a
m }P_b-\delta_{b m }P_a\right)\nonumber\\
&&~~~~~~~~~~~~~~~~~~~~~~~~~~~~+i\left[\,M_{ab},M_{k l
}\,\right]\wedge
\sin(\alpha_5 P_m ) \notag \\
&&~~~~~~~~~~~~~~~~~~~~~~~~+\left[\,\left[\, M_{ab},M_{k l
}\,\right],M_{k l }\,\right]\perp
(\cos(\alpha_5  P_m  )-1) \label{coproduct50} \\
&&~~~~~~~~~~~~~~~~~~~~~~~~+ M_{k l }\sin(\alpha_5 P_m )\perp
\alpha_5 \left(\psi_m P_k -\chi_m P_l \right) \nonumber  \\
&&~~~~~~~~~~~~~~~~~~~~~~~~+ \alpha_5 \left(\psi_m P_l +\chi_m P_k
\right)\wedge M_{k l
}(\cos(\alpha_5 P_m )-1)\;, \nonumber\\
&~~&  \cr &&\Delta_{\alpha_5}(K_a) = \Delta_0(K_a)
-i\left[\,K_a,M_{k l }\,\right]\wedge \sin(\alpha_5 P_m )
\label{coproduct5}\\
&&~~~~~~~~~~~~~~~~~~~~~~~~~~~~~~~~~~~+\left[\,\left[%
K_a,M_{k l }\,\right],K_{k l }\,\right]\perp (\cos(\alpha_5 P_m
)-1)\;,\cr &~~&  \cr &&\Delta_{\alpha_5}(P_a) = \Delta _0(P_a)+\sin(
\alpha_5 P_m )\wedge
\left(\delta_{k a}P_l -\delta_{l a}P_k \right)\\
&&~~~~~~~~~~~~~~~~~~~~~~~~~~~~~~~~~~~+(\cos(\alpha_5  P_m )-1)\perp
\left(\delta_{k a}P_k +\delta_{l a}P_l \right)\;, \cr &~~& \cr
&&\Delta_{\alpha_5}(H) = \Delta_0(H) \pm \frac{\alpha_5}{\tau^2}K_m
\wedge M_{kl} \;,\label{coproduct5a}
\end{eqnarray}
for the last case, where
\begin{eqnarray}
\psi_m = \delta_{b m }\delta_{l a}-\delta_{a m }\delta_{l
b}\;\;\;,\;\;\;\chi_m = \delta_{b m }\delta_{k a}-\delta_{a m
}\delta_{k b}\;\;\;,\;\;\; a \perp
 b = a \otimes b + b\otimes a\;.
\end{eqnarray}

 The algebraic sector (\ref{nnnga}) together with classical
 antipodes and coalgebraic relations (\ref{coproduct1})-(\ref{coproduct5a}) define
the  $\alpha_i$-deformed Newton-Hooke Hopf algebras ${\mathcal
U}_{\alpha_i}({ NH}_{\pm})$ respectively $(i=1,2,...,5)$. It should
be noted that for all parameters $\alpha_i$ running to zero the
above Hopf structures   become classical. Besides, one can also
observe that for  time scale $\tau$ approaching infinity  we get
$\,{\mathcal U}_{\alpha_i}({\mathcal G})$ Galilei quantum groups
constructed in \cite{dasz1}.

\section{{{Quantum  Newton-Hooke space-times}}}

Let us now turn to the deformed space-times corresponding to the
Hopf algebras provided in pervious Section. They are defined as the
quantum representation spaces (Hopf modules) for quantum
Newton-Hooke algebras, with action of the deformed symmetry
generators satisfying suitably deformed Leibnitz rules \cite{bloch},
\cite{wess}, \cite{chi}. \\
The action of Newton-Hooke groups $\,{\mathcal U}_{\alpha_i}({
NH}_{\pm})$ on a Hopf module of functions depending on space-time
coordinates $(t,x_a)$ is given by
\begin{equation}
H\rhd f(t,\overline{x})=i{\partial_t}f(t,\overline{x})\;\;\;,\;\;\;
P_{a}\rhd f(t,\overline{x})=iC_{\pm} \left(\frac{t}{\tau}\right)
{\partial_a}f(t,\overline{x})\;, \label{a1}
\end{equation}
and
\begin{equation}
M_{ab}\rhd f(t,\overline{x}) =i\left( x_{a }{\partial_b} -x_{b
}{\partial_a} \right) f(t,\overline{x})\;\;\;,\;\;\; K_a\rhd
f(t,\overline{x}) =i\tau \,S_{\pm} \left(\frac{t}{\tau}\right)
{\partial_a} \,f(t,\overline{x})\;,\label{dsf}
\end{equation}
with $C_{+} [\frac{t}{\tau}] = \cosh \left[\frac{t}{\tau}\right]$,
$C_{-} [\frac{t}{\tau}] = \cos \left[\frac{t}{\tau}\right]$, $S_{+}
[\frac{t}{\tau}] = \sinh \left[\frac{t}{\tau}\right]$, $S_{-}
[\frac{t}{\tau}] = \sin \left[\frac{t}{\tau}\right]$.\\
Moreover, the $\star$-multiplication of arbitrary two functions  is
defined as follows
\begin{equation}
f(t,\overline{x})\star_{{\cdot}} g(t,\overline{x}):=
\omega\circ\left(
 \mathcal{F}_{\cdot}^{-1}\rhd  f(t,\overline{x})\otimes g(t,\overline{x})\right)
 \;,
\label{star}
\end{equation}
where symbol  $\mathcal{F}_{\cdot}$ denotes the  twist factor
corresponding to a proper Newton-Hooke group and $\omega\circ\left(
a\otimes b\right) = a\cdot b$.

Consequently, we get:\\

{\bf 1)} The Newton-Hooke space-times corresponding to the quantum
groups $\,{\mathcal U}_{\alpha_1}({ NH}_{\pm})$.  Then, the proper
twist factors look as follows
\begin{eqnarray}
\mathcal{F}_{\alpha_1} = \exp
\left(\frac{i}{2}\alpha_1^{kl}\,C_{\pm}^2
\left(\frac{t}{\tau}\right){\partial_l}\wedge {\partial_k}
\right)\;,
\end{eqnarray}
while the nonrelativistic space-times take the form
\begin{eqnarray}
&&[\,t,x_a\,]_{{\star}_{\alpha_1}} =0\;\;\;,\;\;\;
[\,x_a,x_b\,]_{{\star}_{\alpha_1}} =i\alpha_1^{kl}\,C_{\pm}^2
\left(\frac{t}{\tau}\right)(\delta_{ak}\delta_{bl} -
\delta_{al}\delta_{bk})\;.\nonumber\label{spacetime1}
\end{eqnarray}\\

{\bf 2)} The quantum spaces associated with $\,{\mathcal
U}_{\alpha_2}({ NH}_{\pm})$ Hopf algebras. In such a case the twist
factors and corresponding space-times are given by
\begin{eqnarray}
\mathcal{F}_{\alpha_2} = \exp
\left(\frac{i}{2}\alpha_2^{kl}\tau\,C_{\pm}
\left(\frac{t}{\tau}\right)S_{\pm}
\left(\frac{t}{\tau}\right){\partial_l}\wedge {\partial_k}
\right)\;,
\end{eqnarray}
and
\begin{eqnarray}
&&[\,t,x_a\,]_{{\star}_{\alpha_2}} =0\;\;\;,\;\;\;
[\,x_a,x_b\,]_{{\star}_{\alpha_2}} =i\alpha_2^{kl}\tau\, C_{\pm}
\left(\frac{t}{\tau}\right)S_{\pm}
\left(\frac{t}{\tau}\right)(\delta_{ak}\delta_{bl} -
\delta_{al}\delta_{bk})\;,\nonumber\label{spacetime2}
\end{eqnarray}
respectively. \\

{\bf 3)} The deformed space-times associated with $\,{\mathcal
U}_{\alpha_3}({ NH}_{\pm})$ quantum groups. Then, we get the
following twist factors
\begin{eqnarray}
\mathcal{F}_{\alpha_3} = \exp
\left(\frac{i}{2}\alpha_3^{kl}\tau^2\,S_{\pm}^2
\left(\frac{t}{\tau}\right){\partial_l}\wedge {\partial_k}
\right)\;,
\end{eqnarray}
and the proper  Newton-Hooke spaces
\begin{eqnarray}
&&[\,t,x_a\,]_{{\star}_{\alpha_3}} =0\;\;\;,\;\;\;
[\,x_a,x_b\,]_{{\star}_{\alpha_3}} =i\alpha_3^{kl}\tau^2\, S_{\pm}^2
\left(\frac{t}{\tau}\right)(\delta_{ak}\delta_{bl} -
\delta_{al}\delta_{bk})\;.\nonumber\label{spacetime3}
\end{eqnarray}\\

{\bf 4)} The fourth class of space-times  $\,({\mathcal
U}_{\alpha_4}({ NH}_{\pm}))$
\begin{eqnarray}
&&[\,t,x_a\,]_{{\star}_{\alpha_4}} =0\;\;\;,\;\;\;
[\,x_a,x_b\,]_{{\star}_{\alpha_4}} =2i\alpha_4\tau\, S_{\pm}
\left(\frac{t}{\tau}\right)\left[\;\delta_{ma}(x_k\delta_{bl} -
x_{l}\delta_{bk}) - \delta_{mb}(x_k\delta_{al} -
x_{l}\delta_{ak})\;\right]\;,\nonumber\label{spacetime4}
\end{eqnarray}
corresponding to the following twist factors
\begin{eqnarray}
\mathcal{F}_{\alpha_4} = \exp \left(i\alpha_4\tau\,S_{\pm}
\left(\frac{t}{\tau}\right) \left( x_{k }{\partial_l} -x_{l
}{\partial_k} \right)\wedge {\partial_m}\right) \;.
\end{eqnarray}\\

{\bf 5)} The last kind of quantum spaces associated with
$\,{\mathcal U}_{\alpha_5}({ NH}_{\pm})$ groups. In such a case the
twist factors take the form
\begin{eqnarray}
\mathcal{F}_{\alpha_5} = \exp \left(i\alpha_5\,C_{\pm}
\left(\frac{t}{\tau}\right) \left( x_{k }{\partial_l} -x_{l
}{\partial_k} \right)\wedge {\partial_m}\right) \;.
\end{eqnarray}
while the noncommutative space-times look as follows
\begin{eqnarray}
&&[\,t,x_a\,]_{{\star}_{\alpha_5}} =0\;\;\;,\;\;\;
[\,x_a,x_b\,]_{{\star}_{\alpha_5}} =2i\alpha_5\,C_{\pm}
\left(\frac{t}{\tau}\right)\left[\;\delta_{ma}(x_k\delta_{bl} -
x_{l}\delta_{bk}) - \delta_{mb}(x_k\delta_{al} -
x_{l}\delta_{ak})\;\right]\;.\nonumber\label{spacetime5}
\end{eqnarray}\\

Let us note that due to the form of functions $C_{\pm}
(\frac{t}{\tau})$ and $S_{\pm} (\frac{t}{\tau})$ the spatial
noncommutativities {\bf 1)} - {\bf 5)} are expanding or periodic in
time for $\,{\mathcal U}_{\alpha_i}({ NH}_{+})$ and $\,{\mathcal
U}_{\alpha_i}({ NH}_{-})$ Hopf algebras respectively. It should be
also noted that for deformation parameters $\alpha_i$ approaching
zero all above space-times become classical. Besides, as it was
mentioned in Introduction, for 
time parameter $\tau$ running to infinity we recover the Galilei
quantum space-times
provided in \cite{dasz1} and \cite{dasz2}, i.e. we get 
\begin{eqnarray}
&{ \bf i)}&\;\;[\,t,x_a\,]_{{\star}_{\alpha_1}} =0\;\;\;,\;\;\;
[\,x_a,x_b\,]_{{\star}_{\alpha_1}}
=i\alpha_1^{kl}(\delta_{ak}\delta_{bl} - \delta_{al}\delta_{bk})\;,\;\;\;\\
&~~&\cr &{ \bf ii)}&\;\;[\,t,x_a\,]_{{\star}_{\alpha_2}}
=0\;\;\;,\;\;\; [\,x_a,x_b\,]_{{\star}_{\alpha_2}}
=i\alpha_2^{kl}\,t\,(\delta_{ak}\delta_{bl} - \delta_{al}\delta_{bk})\;,\;\;\;\\
&~~&\cr &{ \bf iii)}&\;\;[\,t,x_a\,]_{{\star}_{\alpha_3}}
=0\;\;\;,\;\;\; [\,x_a,x_b\,]_{{\star}_{\alpha_3}}
=i\alpha_3^{kl}\,t^2\, (\delta_{ak}\delta_{bl} -
\delta_{al}\delta_{bk})\;,\;\;\;~~~~~~~~~\\ &~~&\cr &{ \bf
iv)}&\;\;[\,t,x_a\,]_{{\star}_{\alpha_4}} =0\;\;\;,\;\;\;
[\,x_a,x_b\,]_{{\star}_{\alpha_4}} =2i\alpha_4\,t\,
\left[\;\delta_{ma}(x_k\delta_{bl} - x_{l}\delta_{bk}) -
\delta_{mb}(x_k\delta_{al} - x_{l}\delta_{ak})\;\right]\;,\nonumber\\
&~~&\cr &{ \bf v)}&\;\;[\,t,x_i\,]_{{\star}_{\alpha_5}}
=0\;\;\;,\;\;\; [\,x_a,x_b\,]_{{\star}_{\alpha_5}}
=2i\alpha_5\,\left[\;\delta_{ma}(x_k\delta_{bl} - x_{l}\delta_{bk})
- \delta_{mb}(x_k\delta_{al} -
x_{l}\delta_{ak})\;\right]\;,\nonumber
\end{eqnarray}
respectively.  The quantum space ${ \bf i)}$ corresponds to the
canonical type of noncommutativity (\ref{noncomm}), the space-times
${ \bf ii)}$ and ${ \bf v)}$ - to the Lie-algebraic class
(\ref{noncomm1}), while the quantum spaces ${ \bf iii)}$ and ${ \bf
iv)}$ belong to the quadratic type of space-time deformation
(\ref{noncomm2}).

\section{{{Final remarks}}}

In this article we introduce five twisted Newton-Hooke  Hopf
algebras and the  corresponding  deformed space-times with quantum
space and classical time (see {\bf 1)} - {\bf 5)}). We demonstrate
that derived quantum  spaces are respectively periodic and expanding
in time, with  $\,{\mathcal U}_{\alpha_i}({ NH}_{-})$ and
$\,{\mathcal
U}_{\alpha_i}({ NH}_{+})$ quantum groups as symmetries. 
In the limit $\tau \to \infty$ we also recover five twisted Galilei
quantum spaces {\bf i)} - {\bf v)} proposed in \cite{dasz1} and
\cite{dasz2}.

It should be noted that  present studies can be extended in various
ways.  First of all, one can find the dual Hopf structures
$\,{\mathcal D}_{\alpha_i}(NH_{\pm})$ with the use of FRT procedure
\cite{frt} or by canonical quantization of the proper Poisson-Lie
structures \cite{poisson}. Besides, as it was mentioned in
Introduction,  one can look for dynamical models corresponding to
the Newton-Hooke space-times $\bf 1)$ - $\bf 5)$. In the case of
twisted Galilei Hopf algebras such investigations have been
performed in \cite{walczyk1} and \cite{walczyk2}. Finally, one can
consider the more complicated (non-Abelian\footnote{There exist yet
one (omitted in this article) Abelian twist factor with carrier
$M_{ab} \wedge H$. However, due to the form of representation
(\ref{a1}), (\ref{dsf}), it provides well known, less interesting
type of noncommutativity (\ref{noncomm1}).}) twist deformations of
Newton-Hooke Hopf algebras, and find subsequently the twisted
coproducts, corresponding noncommutative space-times and dual
Newton-Hooke  Hopf structures. The mentioned problems are now under
consideration.

\section*{Acknowledgments}
The author would like to thank J. Lukierski and M. Woronowicz
for valuable discussions.\\
This paper has been financially supported by Polish Ministry of
Science and Higher Education grant NN202318534.

\end{document}